%% file: main.tex
\documentclass[conference]{IEEEtran}
\IEEEoverridecommandlockouts
\usepackage{cite}
\usepackage{comment}
\usepackage{amsmath,amssymb,amsfonts}
\usepackage{algorithmic}
\usepackage{multirow}
\usepackage{textcomp}
\usepackage{xcolor}
\usepackage[normalem]{ulem}
\usepackage[switch]{lineno}
\usepackage[utf8]{inputenc}
\usepackage{csquotes}
\usepackage[english]{babel}
\usepackage{array}
\usepackage{float}
\usepackage{enumitem}
\newcolumntype{P}[1]{>{\centering\arraybackslash}p{#1}}
\usepackage{multirow}
\usepackage{graphicx}
\usepackage{comment}
\usepackage{ulem}

\usepackage{subcaption}
\usepackage{url}

\usepackage{breakurl}
\usepackage{hyperref}
\hypersetup{breaklinks}
\usepackage{ulem}
\makeatletter
\usepackage{capt-of}
\def\BibTeX{{\rm B\kern-.05em{\sc i\kern-.025em b}\kern-.08em
    T\kern-.1667em\lower.7ex\hbox{E}\kern-.125emX}}
\begin{document}

\title{Overcoming Pedestrian Blockage in mm-Wave Bands using Ground Reflections 
}
\author{\IEEEauthorblockN{Santosh Ganji, Romil Sonigra, and P. R. Kumar, Fellow, IEEE}\\
\IEEEauthorblockA{Texas A\&M  University}}

 \maketitle

\begin{abstract}
mm-Wave communication employs directional beams to overcome high path loss. High data rate communication is typically along line-of-sight (LoS). In outdoor environments, such communication is susceptible to temporary blockage by pedestrians interposed between the transmitter and receiver. It results in outages in which the user is ``lost," and has to be re-acquired as a new user, severely disrupting interactive and high throughput applications. 
It has been presumed that the solution is to have a densely deployed set of base stations that will allow the mobile to perform a handover to a different non-blocked base station every time a current base station is blocked. 
This is however a very costly solution for outdoor environments.

Through extensive experiments 
we show that it is possible to exploit a strong ground reflection with a received signal strength (RSS) about 4dB less than the LoS path in outdoor built environments with concrete or gravel surfaces, for beams that are narrow in azimuth but wide in zenith.
While such reflected paths cannot support the high data rates of LoS paths, they can 
support control channel communication, and, importantly, sustain time synchronization between the mobile and the base station.
This allows a mobile to quickly recover to the LoS path upon the cessation of the temporary blockage, which typically lasts a few hundred milliseconds.
We present a simple in-band protocol that quickly discovers ground reflected radiation and uses it to sustain link during temporary blockage.

\end{abstract}

\input{Introduction}

\input{related_work}
\input{background}
\input{Measurements}
\input{protocol}
\input{Comparison}
\bibliographystyle{IEEEtran}
\bibliography{IEEEabrv,references}
\end{document}

%% file: Introduction.tex
\section{Introduction}
mm-Wave communication systems use highly directional beams to combat high path loss. The base station and the mobile typically employ their beams in the Line of Sight (LoS) direction to maximize link strength for high data rate communication. Due to high oxygen absorption \cite{NLOSmain1} at mm-wave frequencies from $28$ to $80$ GHz there is an at least $40$ dB absorption loss. Consequently, a human body interposed between the base station and the mobile completely blocks the narrow directional link between mobile and base station.  In outdoors settings, when pedestrians obstruct the LoS path between mobile and base station, 
 packet communication between mobile and base station is disrupted. The mobile is essentially ``lost" to the base station, and will need to be re-acquired as a new user. This can however involve a delay of about a second. 
 

 For example, in 5G New Radio, for a mobile to start the initial access procedure, the base station sweeps 
 broadcast information in $64$ beams every $20$ ms. So even to discover a base station beam, the mobile can take up to 1.28 seconds \cite{_2017_nr2}. This  not only incurs significant power consumption, but it also
disrupts many mobile applications, e.g.,
user experience for high throughput applications like Virtual Reality (VR) and high-definition video streaming.
Blockage by pedestrians is therefore regarded as a serious problem outdoors
\cite{Ishmobile,Pedestrianfading,Pedestrianfading1}.

To avoid such disruption it has been suggested that the mobile can perform handover to a different nearby base station with an LoS beam \cite{Ishmobile}.
This however requires a dense, and therefore costly, deployment of base stations in outdoor environments.
Another suggested solution is Coordinated Multipoint Transmission
\cite{BS1,BS2}, which also requires a dense deployment of base stations.
Prior works have studied cell density requirements \cite{Ishmobile} and the extent of coordination needed between base stations \cite {BS1,BS2,comp} during blockage events. In addition, it incurs delay since to discover a neighboring base station beam and perform handover, the mobile needs to perform cell beam discovery, respond with random access, followed by exchange of control plane messages,
e.g., to perform authentication, etc.
It also consumes power from the battery-driven mobile \cite{HO_Challenges}. 

Avoiding a dense and costly deployment of base stations, and the coordination challenges it presents \cite {BS1,BS2,comp,Ishmobile},
necessitates a different solution for outdoor environments. Through extensive
%
%
%
outdoor signal measurement campaigns using $60$ GHz transceivers, we show that it is possible to exploit a ground reflection that has a signal strength about 4dB less than a LoS path in outdoor environments with hard surfaces such as concrete or gravel, when the beam is narrow in azimuth but wide in zenith.\footnote{{ We have also determined through experimentation that such a ground reflection also exists indoors, from hard surfaces such as ceramic tiles. Such reflections have not apparently been measured indoors previously, possibly because they require antenna array placement at a height and directed at an appropriate angle to allow the reflected beam to reach the mobile in spite of blockage by another human. In fact the reflected beam so generated is better than the RSS on NLoS paths previously measured in indoor environments via walls, which is about 10 dB lower than LoS paths \cite{unblock}.} }  While such reflected beams, { indoor or outdoors, cannot support the high rate of the LoS communication prior to blockage}, they can support lower rate bidirectional   channel communication.
We show that it allows time synchronization between the mobile and the base station to be sustained, so that the mobile can quickly revert to LoS communication as soon as the temporary blockage, which typically lasts a few hundred milliseconds, ends.
%
%
We also present an in-band protocol to discover ground reflections and quickly recover LoS communications after temporary blockages. 
This allows us to extend the BeamSurfer protocol for indoor environments \cite{BeamSurfer} to outdoor environments.

%% file: related_work.tex
\section{Related Work}
The existence of ground reflections is well known. In fact it is the basis of instrumented landing systems \cite{Glide_Scope} operating in the UHF 300 to 100 MHz band wave band { that use a reflected wave to form a glide path for landing}.
For communications, Rajagopal et al \cite{NLOSmain1}
and Jaeckel et al \cite{GRmodel1} have
conducted measurement campaigns in mm-wave bands to characterize ground reflections. Specifically, \cite{NLOSmain1} has identified the presence of strong ground reflections comparable to LoS directions 
at $28$ GHz in outdoor environments, {which from the presented graphs appear to be about
6-8dB below LoS signal strength}. 

Several works \cite{Pedestrianfading,handgrip,HM1,HM2}  have also shown that human body blockage is severe at mm-wave frequencies. The effect of pedestrian traffic on mm-wave systems has been studied in \cite{walk,blockage_walk,unblock}.  In particular, they have characterized the duration of pedestrian blockage events.
It has been observed that blockage lasts for 100 to 300 milliseconds in crowded environments.

To overcome the ill effects of pedestrian blockage, 
researchers have broadly proposed two approaches. One is to directly use NLoS paths from the environment for data communication, and the other is
to deploy a large number of base stations. 





 {The experimental measurement campaign reported in \cite{Rappaport}, specifically mentions
scatterers including ``lamppost, building, tree, or automobile", but not specifically the ground itself, and reports NLoS beams that have 10-50 dB more loss than LoS.
One approach, 
BeamSpy \cite{beamspy}, suggests the employment of a full geometric model
at an anchor location 
through a full space scan. It proposes that a predicted set of transmitter and receiver beam pairs that capture NLoS paths 
at a neighbor location to the anchor point
be tested by the receiver to
identify a useful pair that restores link signal strength during the blockage. 
The geometric model needs be recreated if the anchor point is far away, i.e., more than 3 m, from the neighbor location. 
The work reported in
\cite{SimulationStudy} simulated beam switching from LoS to NLoS beams during blockage, and assert that most NLoS paths from the environment are not useful for high data rate communication.} 

The density of base stations necessary
to meet the performance of high throughput and low latency applications is studied in  \cite{Ishmobile}.  To meet 5G New Radio application requirements 
in outdoor environments, it is determined that a base station density of 200 $BS/km^2$ is needed. 
Such a high base station density demands tight network coordination to manage inter-cell interference, and \cite{BS1} analyzed coordinated multiple access to switch base stations during blockage events, and mention 
an inverse relation at high base station density between reliable blockage recovery and throughput. 


Our experiments, conducted in the $60$ GHz band, show that strong ground reflections from {multiple surfaces outdoors by surfaces such as concrete and gravel, and indoors via hard surfaces such as ceramic tile,}  can be used to overcome pedestrian blockage problems by sustaining time synchronization and control channels during blockage, and allowing quick recovery to a high data rate LoS when blockage disappears. 

%% file: background.tex
\section{Background} \label{background}

Cellular networks rely on  mm-wave spectrum to provide gigabit throughput utilizing the large bandwidth available. Due to their smaller wavelengths, mm-wave links have high path and penetration losses. To overcome link losses in mm-wave networks, the base station and the mobile devices communicate in a directional fashion using narrow radio beams. While the power amplifiers in the radio front end provide certain gains, using the directivity gain from a passive antenna array is critical for mobile mm-wave devices. For minimal link loss, the base station and mobile need to communicate using a directional beam in the LoS direction.

In the sub-6 GHz frequencies, the environment scatters electromagnetic radiation from an omni-directional transmitter in all directions. An omni-directional receiver can capture this incoming radiation i.e., all the multi-path components of the transmitted signal. In contrast, due to directional transmission, there are fewer and more distinct multi-path components in the mm-wave bands. A narrow directional radio receiver beam can only receive signal components that arrive in the beam direction.  To discover either LoS or NLoS paths, the base station sweeps beams within a sector and the  mobile receiver performs a spatial scan.   NLoS signal components in mm-wave have (RSS) at least 10 dB less than that of LoS paths. The receiver may discover an LoS path using wider beams but it cannot discover NLoS paths as they have much lower signal strength.

In the sub 6-GHz band, an obstacle interposed between the base station and the mobile cannot block the transmission
as the receiver can capture a large number of multipath components. However, an interposed pedestrian does obstruct a directional mm-wave  LoS link.
The poor RSS during blockage events leads to link outage. The mobile is left with one of two choices to continue communication with the network -- to switch to a NLoS path if such a path exists between the base station and mobile, or to perform handover to a neighboring base station (or otherwise employ a neighboring base station through, say, coordinated multipoint transmission). NLoS paths between the base station and the mobile typically exist in indoor environments.

The base station and mobile need to adapt their LoS beams to compensate for user mobility. They use phased antenna arrays that electronically steer the direction of radio beams. As the size of the array increases i.e., as the number of antenna elements increases, the directivity gain increases, and the resulting radio beam has a smaller beamwidth. For example, a 32x32 element uniform planar array can produce beamwidths as narrow as $4^\circ$. Using these narrow directional beams, the mobile must scan the full space to discover a NLoS path.  Typically, both the mobile and the base station use beam codebooks that have pre-calculated phase weights { that yield beams in specific directions}. Employing these weights, the array steers the beam in desired directions. The mobile and the base station use beams from their respective codebooks both to communicate along the LoS path as well as to discover NLoS paths. The number of measurements required to identify at least one NLoS path is proportional to the number of beams in the codebook.

Performing a complete environment scan is not only resource intensive but it also entails usage of a number of signal measurement opportunities. For the mobile to scan the environment, the base station has to allocate measurement  opportunities while catering to data demands of entire network. The mobile needs  measurement schedules until it discovers at least one NLoS path. The number of measurements depends on angular resolution of spatial scan. 

 Link blockage, especially by pedestrians is a sudden and unpredictable event.
The mobile must always have in hand at least one NLoS direction, that it can use to avoid an outage when a blockage occurs. 
As the user moves, the environment changes, so the mobile needs to perform frequent environment scans.

For optimal link performance, both the base station and the mobile must continually adapt their respective directional radio beams to maintain highly aligned LoS beams. During user mobility, beam adaptation is quite challenging and requires several measurement opportunities. Also, to discover NLoS paths, the mobile needs significant additional measurement opportunities. Every time the base station adapts the transmit beam to counter user mobility, the mobile needds to perform a full spatial scan to discover new NLoS paths. As transmit beam adaptation happens frequently during user mobility, the NLoS path discovery process is performed often. The frequent ambient scans reduce the overall network throughput as the base station is responsible for scheduling measurement opportunities for all the users. 

If the mobile does not have an NLoS path in its memory during a blockage event, then link outage occurs, and the mobile gets disconnected from the base station. The mobile will then need to perform an initial network access procedure just as though it were a new user. Similarly, the mobile will also need to perform a similar procedure to handover to a neighboring base station. Base stations periodically sweep directional beams with reference signals and 
broadcast information such as cell and network identity. A mobile sweeps through all its receive beams one at a time to discover at least one of the base station's beams. To complete bi-directional connection, the mobile transmits a random preamble in the same direction in which it discovered the base station's beam, and awaits a response. After physical layer procedures to establish reliable data communication, the network then authenticates the mobile before granting network access. This complete procedure takes several seconds to complete handover 
\cite{HO_Challenges}.

In indoor environments, \cite{unblock,Ish,beamspy,BeamSurfer} have suggested harvesting NLoS paths to preserve the link between the base station and the mobile during transient blockage events. For outdoor environments, in contrast, dense base station deployment and switching base stations in case of blockage has been suggested \cite{Ish,Relay_assisted}.

\begin{figure}[h]
  \centering
 \includegraphics[width=.8\linewidth, height=1.5in]{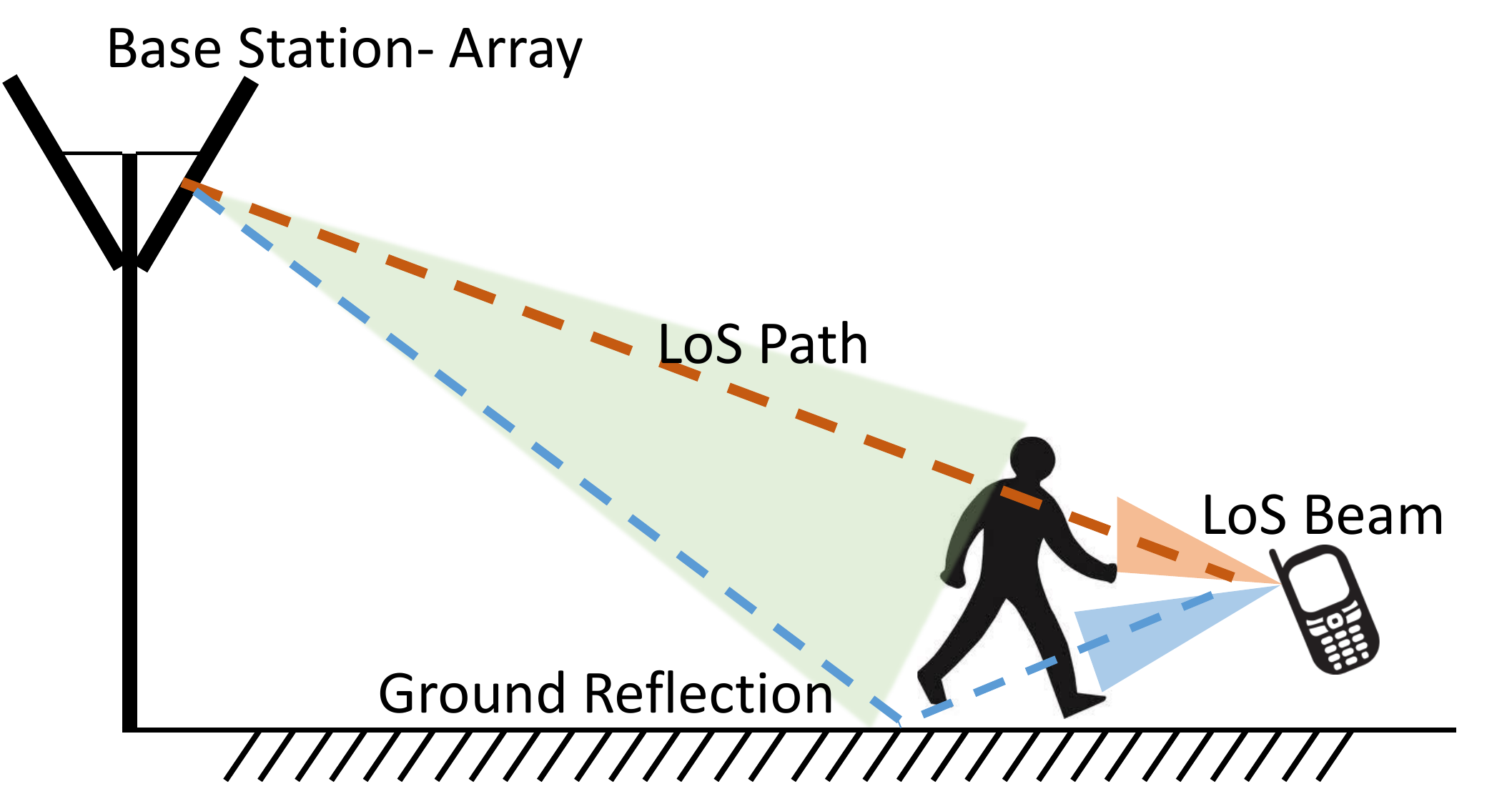}
  \caption{Ground Reflection }
  \label{GR}
\end{figure}

However, outdoors too, there can be strong ground reflections. In fact, such reflections are used to shape glide paths for instrument landing systems for aircraft \cite{Glide_Scope}.
Outdoor reflections have also been investigated for communication \cite{NLOSmain1,GRmodel1}. Motivated by this possibility, we have conducted extensive experiments in the 60GHz band, and have observed that mm-wave signals are reflected from outdoor surfaces such as concrete and gravel.
As base stations are usually deployed with a slight downward tilt, as shown in Figure \ref{GR}, and are equipped with phased arrays that steer beams, the mobile's receiver can capture these ground reflections. The ground reflections can be found in the same azimuth direction as the LoS path. To investigate if such reflections are usable during blockage events, we performed link measurements with human's blocking the LoS link between the base station and the mobile.  We have found that  
even under the presence of a human blocker,
there is a ground reflection with an RSS that is within 6 dB of the RSS of a direct unblocked LoS link. We have repeated the experiments with different surfaces, and have
found that in some scenarios, ground reflections are even strong enough to handle 
limited data plane traffic. Our experiments and observations are detailed in Section \ref{measurements}. 

%% file: measurements.tex

\section{Measurements} \label{measurements}

To measure the signal strength of ground reflections at mm-wave frequencies, in particular at 60 GHz, we performed measurements in environments with commonly found ground surfaces. 
We conducted experiments using software-defined radios operating at 60 GHz \cite{instruments_2020_introduction}. Baseband IQ sample generation at transmitter and signal processing at the receiver are implemented in FPGA. An analog baseband signal of 2 GHz bandwidth is upconverted to 60 GHz carrier frequency. A 12 element phased array is used both at the transmitter and receiver.

\begin{figure}
	\centering
	\begin{minipage}{.45\columnwidth}
		\centering
		\includegraphics[width=1\textwidth]{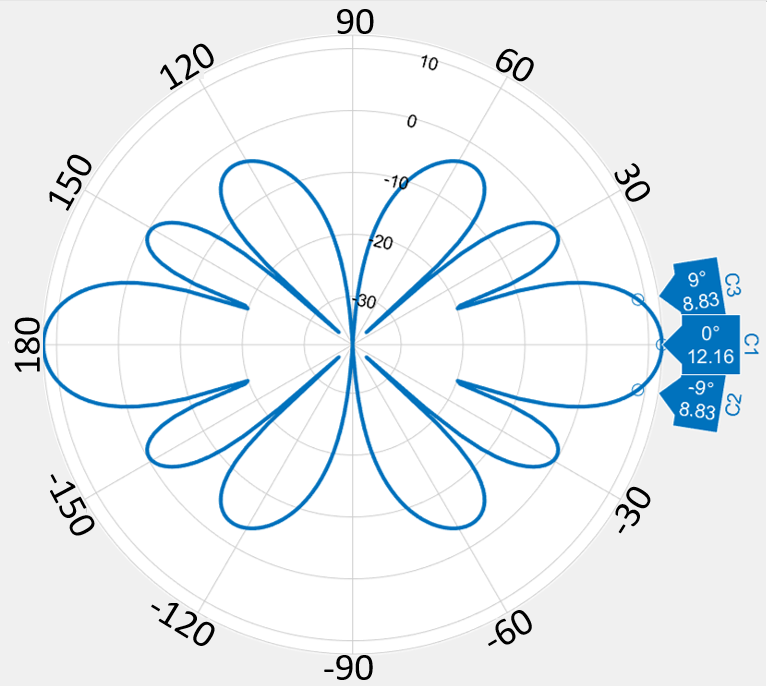}
		\caption{Azimuth Cut}\label{Azimuth}
	\end{minipage}%
	\hfill
	\begin{minipage}{.45\columnwidth}
		\centering
		\includegraphics[width=\textwidth]{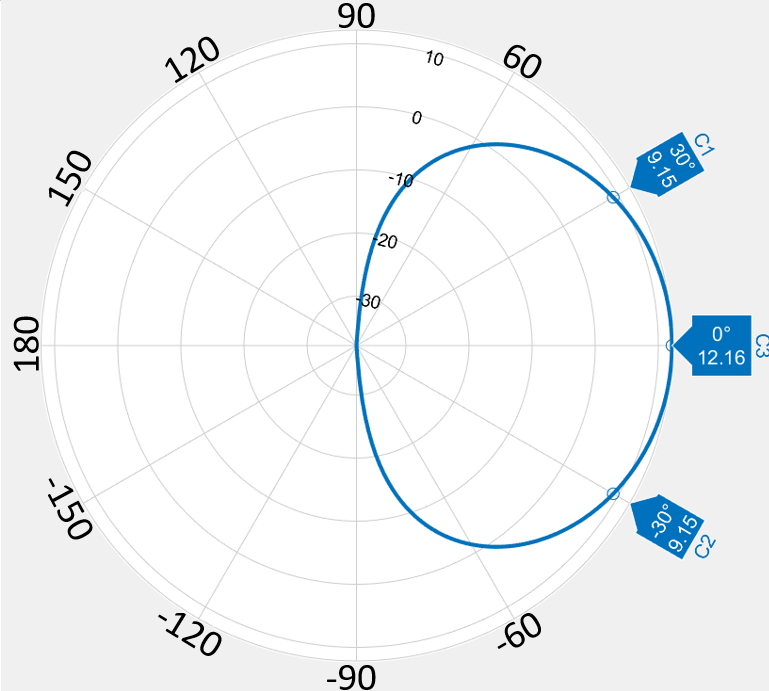}
		\caption{Elevation Cut}\label{Elevation}
	\end{minipage}
\end{figure}

The phase weights for desired radiation patterns are calculated and stored as beam codebooks.  
Our beam codebook has 25 beams, with narrow beams 
of width approximately 
$18^\circ$,
within a $120^\circ$ azimuth sector. Further details on the transceiver design and implementation are available in \cite{unblock,x60}. Figs. \ref{Azimuth} and \ref{Elevation} present Azimuth and Elevation radiation patterns of the bore sight beam. The zenith beamwidth is around $60^\circ$, whereas azimuth beamwidth is, as noted above, $18^\circ$.  Transmit power is fixed at 20 dBm. The directivity gain of the phased array is 17 dB.

On each surface under study, we set up the transmitter array  2.5 m above ground level using a tripod, with the receiver antenna array held about 1 m from the surface. Transmitter and receiver arrays are positioned facing 
each other, and are placed 6 m apart. For each scenario, we repeated experiments for two different cases, where the Transmitter antenna is tilted towards the ground by $10^\circ$ or $20^\circ$.  This geometry mimics potential outdoor deployments where base stations are located higher than mobiles. This tilt is responsible for creating additional reflected directions towards the receiver. Moreover, most of the elevation beamwidth is directed towards the receiver.


While the transmitter beam is in the LoS direction of the receiver, signal strength at the receiver is measured using a beam that is highly aligned with the transmitter beam. We use $RSS_{LoS}$ to represent the signal strength in the LoS direction at the receiver. It serves as a reference to calculate total loss suffered by ground reflection. $RSS_{LoS}$ in our experiments is $-60$ dBm. When a human obstructs the LoS direction by standing in between transmitter and receiver, the RSS is $-78$ dBm, which is the noise floor of our receiver. This indicates that a pedestrian can completely block a signal. Although the pedestrian blockage is transient, an undesirable outage event occurs at the receiver.

Let $H_T$ and $H_R$ be the distances from ground level to transmitter array and receiver array, respectively. $D_{TR}$ denotes the distance between transmitter and receiver. $H_{B}$ is the height of a human blocker. The blocker can only obstruct the transmission only when she is close to the receiver. Using ray tracing, we can derive the maximum distance between blocker and the receiver $D_{BR_{max}}$ to obstruct LoS transmissions as 
\begin{equation} \label{eqn1}
	D_{BR_{max}} = D_{TR}*\frac{H_{B}-H_R}{H_T-H_R} .
	\end{equation}
For $H_R=7.8$ m of 1.78 m, and $D_{TR}=6$ m, $D_{BR_{max}}$ was found to be 3.12m in our experiments.

\begin{table}[h]
\small 
 \caption{Indoor Floor, Concrete Tiles}
   \label{Indoor}
\begin{center}
\begin{tabular}{ | c | c | c| c|}
\hline
\hline
 Transmitter Tilt  & $RSS_{GR}$ (dBm) & $\phi_{r}$ & $D_{BR}$ (m) \\ \hline \hline
$0^\circ$ & -65.7  & $30^\circ$ & 2\\
$0^\circ$ & -66  & $30^\circ$ & 3\\  
\hline
$10^\circ$ & -64.5  & $9^\circ$ & 2\\  
$10^\circ$ & -64.45  & $10^\circ$ & 3\\
\hline
$20^\circ$ & -64.4  & $30^\circ$ & 2\\ 
$20^\circ$ & -64.3  & $30^\circ$ & 3\\ 
 \hline
\end{tabular}
\end{center}

\end{table}

\begin{table}[h]
\small 
 \caption{Concrete Surface}
   \label{Concrete}
\begin{center}
\begin{tabular}{ | c | c | c| c|}
\hline
\hline
 Transmitter Tilt  & $RSS_{GR}$ (dBm) & $\phi_{r}$ & $D_{BR}$ (m) \\ \hline \hline
$0^\circ$ & -66  & $30^\circ$ & 2\\
$0^\circ$ & -66  & $30^\circ$ & 3\\  
\hline
$10^\circ$ & -64.7  & $14^\circ$ & 2\\  
$10^\circ$ & -64.5  & $18^\circ$ & 3\\  
\hline
$20^\circ$ & -64.1  & $30^\circ$ & 2\\ 
$20^\circ$ & -64  & $30^\circ$ & 3\\ 
 \hline
\end{tabular}
\end{center}

\end{table}

\begin{table}[h]
\small 
 \caption{Gravel Surface}
   \label{Gravel}
\begin{center}
\begin{tabular}{ | c | c | c| c|}
\hline
\hline
 Transmitter Tilt  & $RSS_{GR}$ (dBm) & $\phi_{r}$ & $D_{BR}$ (m) \\ \hline \hline
$0^\circ$ & -66.1  & $30^\circ$ & 2\\
$0^\circ$ & -65.9  & $30^\circ$ & 3\\  
\hline
$10^\circ$ & -64.8  & $19^\circ$ & 2\\  
$10^\circ$ & -64.4  & $15^\circ$ & 3\\  
\hline
$20^\circ$ & -64.4  & $30^\circ$ & 2\\ 
$20^\circ$ & -64.3  & $30^\circ$ & 3\\ 
 \hline
\end{tabular}
\end{center}

\end{table}

Table \ref{Indoor} presents $RSS_{GR}$ averaged over 100 measurements from an indoor surface with concrete tiles. Tables \ref{Concrete} and \ref{Gravel} show RSS from reflections outdoors from  Concrete and Gravel pathways.

When both transmitter and receiver phased arrays are parallel to the ground surface, the only ground reflection available to the receiver is from radiation in one-half of the elevation beamwidth of the transmitter beam. In our case, the elevation beamwidth is $60^\circ$, so the radiation in the bottom half beamwidth reflects from the ground. To capture the reflection in this position, the receiver needs to tilt 
its beams towards the ground while maintaining LoS in azimuth. We observed slightly less $RSS_{GR}$ in this position compared to the positions where the transmitter array is tilted downwards, as the gain of the beam pattern reduces towards the edge of the beamwidth. 

When the transmitter array is tilted towards the ground, directions with stronger incident radiation get reflected, resulting in higher RSS. The highest $RSS_{GR}$ observed on all the 3 surfaces under study is around -64 dBm. This implies that ground reflected radiation is just 4 dB less than LoS. $RSS_{GR}$ is at least 6 dB higher RSS than the NLoS paths \cite{unblock,Ish_two_beams}. It is also important to note that ground reflected radiation takes a shorter path to reach the receiver than NLoS paths from the environment. 

Based on the experiments, the following are our main observations:
\begin{itemize}
    \item Pedestrian blockers can create mm-wave link outage.
    \item Strong ground reflections are available in
    outdoor environments.
    \item Ground reflections are available in the same azimuth LoS direction at the receiver.
      \item Tilting the transmitter towards the ground helps the receiver with even stronger ground reflections. 
      \item Finally, and the most important one, there is no need to handover to z neighboring base station in outdoor environments during transient blockage events.
      
\end{itemize}

%% file: protocol.tex
\section{Protocol to Overcome Pedestrian Blockage}\label{protocol}
From our measurement experiments, we observe that pedestrian blockers block the mm-wave transmissions completely, resulting in link outages. To avoid need for network reconnection, mm-waves devices must remain connected to the network during blockage events. Prior works \cite{beamspy,unblock,BeamSurfer} have identified that by adapting both transmitter and receiver beams in the direction of NLoS paths, devices can continue to communicate during the blockage. However, due to low RSS on NLoS beams, only control plane information can be exchanged reliably. Also, there are several challenges to identify NLoS paths as mentioned in Section \ref{background}.  Our experiments indicate that ground reflections indeed provide very good RSS, so mobiles can even continue a limited amount of data plane traffic during LoS blockage. 

\begin{figure}[h]
  \centering
 \includegraphics[width=.85\linewidth, height=1.5in]{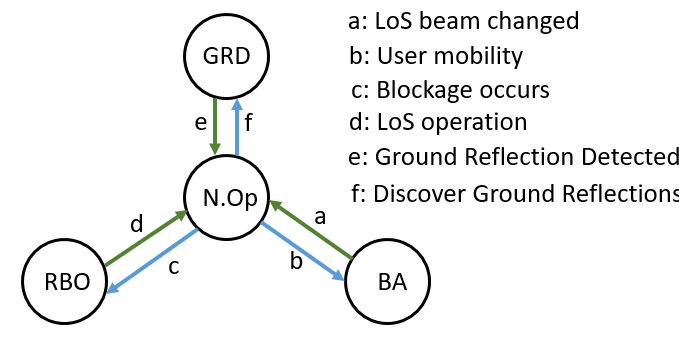}
  \caption{Protocol State Machine }
  \label{SM}
\end{figure}

Based on our observations in Section \ref{measurements}, we present a protocol that mobile devices can follow to recover link signal strength using ground reflections. The protocol is simple and only requires mobiles to make link strength measurements to identify the direction of ground reflections. The protocol is environment agnostic as it relies only on in-band information, specifically RSS. The state machine of the protocol is presented in Fig. \ref{SM}, and is explained below.

During the initial access procedure in directional cellular networks in 5G New Radio, first the mobile discovers the base station beams by sweeping through all its receive beams. It chooses the beam with the highest RSS, and sends a random preamble. The beam with the highest RSS usually is in the line of sight with the base station. The mobile then performs a full spatial scan using its beams. To discover the LoS beam, it may choose to perform either exhaustive search or hierarchical \cite{Agilelink} or compressive sensing based approaches \cite{FALP}. 
We denote the state where the mobile performs initial access as IA. In IA, the mobile discovers the LoS beam to communicate with the base station.

By transmitting random access preamble in the same direction as discovered in the base station beam, the mobile implicitly informs the base station which LoS beam to use to continue communication. Subsequently, as the mobile moves, the base station and the mobile adapt their respective LoS beams to counter user mobility. This adaptation process is called beam alignment, and several protocols \cite{BeamSurfer,Agilelink,beamhistory} have been proposed to adapt the base station and mobile beams during user mobility. Both base station and mobile switch their beams at appropriate times and maintain LoS beams throughout the user mobility. In the beam adaptation state (BA), the mobile continues to maintain an LoS beam during mobility.

Pedestrians can suddenly block the LoS link, in which case RSS suddenly drops \cite{unblock,pedestrians}.   At this point, with our protocol, neither NLoS path discovery nor handover is required at mobile.

We observe from the experiments that the mobile needs to switch only elevation angle to receive ground reflections, while the azimuth direction remains in the LoS direction.   Before the blockage event, the BA state ensures that the mobile is communicating with the base station using LoS beams. Communication using LoS beams is necessary to receive the highest possible signal strength and reach optimal link throughput. Any good beam alignment algorithm \cite{BeamSurfer,Agilelink,beamhistory} ensures that the base station and mobile will use LoS beams to communicate.

As observed from our experiments, the beam that captures the ground reflections is a neighbor beam to LoS beam.  Therefore a mobile connected to the base station needs to switch its current receive beam only in elevation angle to discover the presence of ground reflections.

Suppose that for a base station transmit beam $B_{T_L}$, a highly aligned LoS beam of  mobile is $B_{R_L}$, and the mobile  receives ground reflection using beam $B_{GR}$. To receive ground reflection, the mobile need not perform beam scan in azimuth direction. The azimuth Direction of $B_{GR}$ is in the same direction of $B_{R_L}$. 
Also, the ground reflected path between base station and mobile can be received a by neighboring beam of $B_{R_L}$ i.e., the mobile can identify $B_{GR}$ by switching receive beams to a upward or downward neighbor of $B_{R_L}$.
In the signal domain, the orientation of mobile is not available; therefore, it is not possible to identify whether a beam's upward or downward neighbor $B_{R_L}$ can receive ground reflections. So the mobile makes measurements on both upward and downward neighbors. Let $\Theta_T$ be the tilt of the base station, and $\phi_T$ the elevation beamwidth of $B_{T_L}$. 
Using a ray tracing approach, we can show that the elevation angle of $B_{GR}$ is  $\Theta_T$ + $\frac{\phi_T}{2}$ away from elevation of angle $B_{R_L}$. For example, given $\Theta_T =0^\circ$ and $\phi_T =30^\circ$, the mobile needs to search neighbors within $30^\circ$ of $B_{R_L}$. This search is performed in the state  Ground Reflection Discovery (GRD). Once the $B_{R_L}$  is discovered, the mobile stores it in its memory and uses it in Reflected Beam Operation (RBO) state. The mobile continues to operate in Normal operation state (N.Op) after connecting to the base station until unless beam adaptation  becomes necessary. Fig. \ref{SM}  shows the state machine of the protocol. 

As an illustration of the functioning of the protocol, the following is one plausible sequence of state changes in the protocol: 
\begin{itemize}
    \item Mobile in IA state discovers a base station beam, initiates random access, and establishes network connection. 
    \item It continues to operate in N.Op state till a beam adaptation is necessary.
    \item To recover from blockage, mobile needs knowledge of $B_{GR}$. So it moves to state GRD  at some point during normal operation state to discover ground reflection, and stores $B_{GR}$ in its memory.
    \item In case of blockage, the mobile's state changes to RBO, where it employs $B_{GR}$ to continue communication with base station. After about 100 milliseconds in RBO, it switches to N.Op to see if the LoS beam is free from obstruction.
    \item If the LoS 
    is again available for usage, it remains in N.Op state. 
%
\item As the user moves, the mobile requires LoS beam adaptation, so it moves to state BA.
  
   \item After adapting the LoS beam, the mobile returns to state N.Op.
\end{itemize}

%% file: Comparison.tex
\section{Performance Comparison}
We compare our protocol with two other works, BeamSpy \cite{beamspy} and Unblock \cite{unblock}, which have implemented their respective protocols on 60 GHz testbeds. We implemented Unblock and our protocol on the testbed described in Section \ref{measurements}. For Beamspy \cite{beamspy}, we simply quote its results, since a quasi-omni-directional transmitter 
to which we have no access
is required to reproduce its results.  Our protocol needed 3 measurements to identify ground reflection: first, the mobile measures signal strength on its immediate
neighboring elevation beam to its current receive beam, and then makes the final measurement using the neighboring beam to { then choose the}  beam with higher signal strength from the previous measurements. Table \ref{Performance}  comments on the number of measurements required by each method to discover an NLoS path, as well as the underlying algorithmic complexity. The RSS from ground reflection is 6 dB more than employing a beam discovered by Unblock \cite{unblock} in the indoor environment.

\begin{table}[H]
\small 
 \caption{Performance Comparison}
   \label{Performance}
\begin{center}
\begin{tabular}{ |p{1.5cm}|p{2cm}|p{3cm}| }
\hline
\hline
 Method  &  Number of Measurements & Complexity \\ \hline 
Unblock & 25 & Exhaustive Search\\ \hline 
BeamSpy & 25 & Search followed by offline model creation\\ \hline 
Ground Reflection & 3 & Neighbor Search\\
 \hline
\end{tabular}
\end{center}
\end{table}

\section{Conclusion}
In this work, we have demonstrated that ground reflections can rescue mm-wave links during random and unpredictable pedestrian blockage events. We have presented a simple in-band protocol that mobiles can use to discover ground reflected radiation and recover from temporary pedestrian blockages without outage.

\section*{Acknowledgment}
 This material is based upon work partially supported by US Department of Homeland Security under 70RSAT20CB0000017;
US Office of Naval Research under N00014-21-1-2385, N00014-18-1-2048;
US Army Research Office under W911NF-18-10331, W911NF-21-20064;
US Army Research Lab under W911NF-19-20243;
US National Science Foundation under OMA-2037890, Science $\&$ Technology Center Grant CCF-0939370, CNS-1719384.
The views expressed herein and conclusions contained in this document are those of the authors and should not be interpreted as representing the views or official policies, either expressed or implied, of the U.S. DHS, ONR, ARO, ARL, NSF,  or the United States Government.
The U.S. Government is authorized to reproduce and distribute reprints for Government purposes notwithstanding any copyright notation herein.

%% file: main.bbl
\begin{thebibliography}{10}
\providecommand{\url}[1]{#1}
\csname url@samestyle\endcsname
\providecommand{\newblock}{\relax}
\providecommand{\bibinfo}[2]{#2}
\providecommand{\BIBentrySTDinterwordspacing}{\spaceskip=0pt\relax}
\providecommand{\BIBentryALTinterwordstretchfactor}{4}
\providecommand{\BIBentryALTinterwordspacing}{\spaceskip=\fontdimen2\font plus
\BIBentryALTinterwordstretchfactor\fontdimen3\font minus
  \fontdimen4\font\relax}
\providecommand{\BIBforeignlanguage}[2]{{%
\expandafter\ifx\csname l@#1\endcsname\relax
\typeout{** WARNING: IEEEtran.bst: No hyphenation pattern has been}%
\typeout{** loaded for the language `#1'. Using the pattern for}%
\typeout{** the default language instead.}%
\else
\language=\csname l@#1\endcsname
\fi
#2}}
\providecommand{\BIBdecl}{\relax}
\BIBdecl

\bibitem{NLOSmain1}
S.~Rajagopal, S.~Abu-Surra, and M.~Malmirchegini, ``Channel feasibility for
  outdoor non-line-of-sight mmwave mobile communication,'' in \emph{2012 IEEE
  Vehicular Technology Conference (VTC Fall)}, 2012, pp. 1--6.

\bibitem{_2017_nr2}
\BIBentryALTinterwordspacing
3GPP, ``Nr; physical layer procedures for control,'' 2017. [Online]. Available:
  \url{https://portal.3gpp.org/desktopmodules/Specifications/SpecificationDetails.aspx?specificationId=3215}
\BIBentrySTDinterwordspacing

\bibitem{Ishmobile}
I.~K. Jain, R.~Kumar, and S.~S. Panwar, ``The impact of mobile blockers on
  millimeter wave cellular systems,'' \emph{IEEE Journal on Selected Areas in
  Communications}, vol.~37, no.~4, pp. 854--868, 2019.

\bibitem{Pedestrianfading}
G.~R. MacCartney, T.~S. Rappaport, and S.~Rangan, ``Rapid fading due to human
  blockage in pedestrian crowds at 5g millimeter-wave frequencies,'' in
  \emph{GLOBECOM 2017 - 2017 IEEE Global Communications Conference}, 2017, pp.
  1--7.

\bibitem{Pedestrianfading1}
M.~Abouelseoud and G.~Charlton, ``The effect of human blockage on the
  performance of millimeter-wave access link for outdoor coverage,'' in
  \emph{2013 IEEE 77th Vehicular Technology Conference (VTC Spring)}, 2013, pp.
  1--5.

\bibitem{BS1}
D.~Kumar, J.~Kaleva, and A.~Tölli, ``Blockage-aware reliable mmwave access via
  coordinated multi-point connectivity,'' \emph{IEEE Transactions on Wireless
  Communications}, vol.~20, no.~7, pp. 4238--4252, 2021.

\bibitem{BS2}
D.~Maamari, N.~Devroye, and D.~Tuninetti, ``Coverage in mmwave cellular
  networks with base station co-operation,'' \emph{IEEE Transactions on
  Wireless Communications}, vol.~15, no.~4, pp. 2981--2994, 2016.

\bibitem{comp}
G.~R. MacCartney and T.~S. Rappaport, ``Millimeter-wave base station diversity
  for 5g coordinated multipoint (comp) applications,'' \emph{IEEE Transactions
  on Wireless Communications}, vol.~18, no.~7, pp. 3395--3410, 2019.

\bibitem{HO_Challenges}
M.~Tayyab, X.~Gelabert, and R.~Jäntti, ``A survey on handover management: From
  lte to nr,'' \emph{IEEE Access}, vol.~7, pp. 118\,907--118\,930, 2019.

\bibitem{unblock}
V.~S.~S. Ganji, T.-H. Lin, F.~A. Espinal, and P.~R. Kumar, ``Unblock: Low
  complexity transient blockage recovery for mobile mm-wave devices,'' in
  \emph{2021 International Conference on COMmunication Systems NETworkS
  (COMSNETS)}, 2021, pp. 501--508.

\bibitem{BeamSurfer}
\BIBentryALTinterwordspacing
------, ``Beamsurfer: Simple in-band beam management for mobile mm-wave
  devices,'' in \emph{Proceedings of the SIGCOMM '20 Poster and Demo Sessions},
  ser. SIGCOMM '20.\hskip 1em plus 0.5em minus 0.4em\relax New York, NY, USA:
  Association for Computing Machinery, 2020, p. 15–17. [Online]. Available:
  \url{https://doi.org/10.1145/3405837.3411374}
\BIBentrySTDinterwordspacing

\bibitem{Glide_Scope}
F.~W. Iden, ``Glide-slope antenna arrays for use under adverse siting
  conditions,'' \emph{IRE Transactions on Aeronautical and Navigational
  Electronics}, vol. ANE-6, no.~2, pp. 100--111, 1959.

\bibitem{GRmodel1}
S.~Jaeckel, L.~Raschkowski, S.~Wu, L.~Thiele, and W.~Keusgen, ``An explicit
  ground reflection model for mm-wave channels,'' in \emph{2017 IEEE Wireless
  Communications and Networking Conference Workshops (WCNCW)}, 2017, pp. 1--5.

\bibitem{handgrip}
V.~Raghavan, S.~Noimanivone, S.~K. Rho, B.~Farin, P.~Connor, R.~A. Motos, Y.-C.
  Ou, K.~Ravid, M.~A. Tassoudji, O.~H. Koymen, and J.~Li, ``Hand and body
  blockage measurements with form-factor user equipment at 28 ghz,'' \emph{IEEE
  Transactions on Antennas and Propagation}, pp. 1--1, 2021.

\bibitem{HM1}
M.~Peter, M.~Wisotzki, M.~Raceala-Motoc, W.~Keusgen, R.~Felbecker, M.~Jacob,
  S.~Priebe, and T.~Kürner, ``Analyzing human body shadowing at 60 ghz:
  Systematic wideband mimo measurements and modeling approaches,'' in
  \emph{2012 6th European Conference on Antennas and Propagation (EUCAP)},
  2012, pp. 468--472.

\bibitem{HM2}
T.~Wu, T.~S. Rappaport, and C.~M. Collins, ``The human body and millimeter-wave
  wireless communication systems: Interactions and implications,'' in
  \emph{2015 IEEE International Conference on Communications (ICC)}, 2015, pp.
  2423--2429.

\bibitem{walk}
S.~Collonge, G.~Zaharia, and G.~Zein, ``Influence of the human activity on
  wide-band characteristics of the 60 ghz indoor radio channel,'' \emph{IEEE
  Transactions on Wireless Communications}, vol.~3, no.~6, pp. 2396--2406,
  2004.

\bibitem{blockage_walk}
\BIBentryALTinterwordspacing
A.~Ichkov, P.~M\"{a}h\"{o}nen, and L.~Simi\'{c}, ``End-to-end millimeter-wave
  network performance and mobility management overhead in urban cellular
  deployments with realistic pedestrian traffic and blockages,'' in
  \emph{Proceedings of the 18th ACM Symposium on Mobility Management and
  Wireless Access}, ser. MobiWac '20.\hskip 1em plus 0.5em minus 0.4em\relax
  New York, NY, USA: Association for Computing Machinery, 2020, p. 1–10.
  [Online]. Available: \url{https://doi.org/10.1145/3416012.3424621}
\BIBentrySTDinterwordspacing

\bibitem{Rappaport}
T.~S. Rappaport, Y.~Qiao, J.~I. Tamir, J.~N. Murdock, and E.~Ben-Dor,
  ``Cellular broadband millimeter wave propagation and angle of arrival for
  adaptive beam steering systems (invited paper),'' in \emph{2012 IEEE Radio
  and Wireless Symposium}, 2012, pp. 151--154.

\bibitem{beamspy}
S.~Sur, X.~Zhang, P.~Ramanathan, and R.~Chandra, ``Beamspy: Enabling robust 60
  ghz links under blockage,'' in \emph{Proceedings of the 13th Usenix
  Conference on Networked Systems Design and Implementation}, ser.
  NSDI’16.\hskip 1em plus 0.5em minus 0.4em\relax USA: USENIX Association,
  2016, p. 193–206.

\bibitem{SimulationStudy}
X.~An, C.-S. Sum, R.~V. Prasad, J.~Wang, Z.~Lan, J.~Wang, R.~Hekmat, H.~Harada,
  and I.~Niemegeers, ``Beam switching support to resolve link-blockage problem
  in 60 ghz wpans,'' in \emph{2009 IEEE 20th International Symposium on
  Personal, Indoor and Mobile Radio Communications}, 2009, pp. 390--394.

\bibitem{Ish}
I.~K. Jain, R.~Kumar, and S.~Panwar, ``Driven by capacity or blockage? a
  millimeter wave blockage analysis,'' in \emph{2018 30th International
  Teletraffic Congress (ITC 30)}, vol.~01, 2018, pp. 153--159.

\bibitem{Relay_assisted}
Y.~Liu, Q.~Hu, and D.~M. Blough, ``Blockage avoidance in relay paths for
  roadside mmwave backhaul networks,'' in \emph{2018 IEEE 29th Annual
  International Symposium on Personal, Indoor and Mobile Radio Communications
  (PIMRC)}, 2018, pp. 1--7.

\bibitem{instruments_2020_introduction}
\BIBentryALTinterwordspacing
``Introduction to the ni mmwave transceiver system hardware - national
  instruments,'' 02 2020. [Online]. Available:
  \url{https://www.ni.com/en-us/innovations/white-papers/16/introduction-to-the-ni-mmwave-transceiver-system-hardware.html}
\BIBentrySTDinterwordspacing

\bibitem{x60}
\BIBentryALTinterwordspacing
S.~K. Saha, Y.~Ghasempour, M.~K. Haider, T.~Siddiqui, P.~{De Melo},
  N.~Somanchi, L.~Zakrajsek, A.~Singh, R.~Shyamsunder, O.~Torres, D.~Uvaydov,
  J.~M. Jornet, E.~Knightly, D.~Koutsonikolas, D.~Pados, Z.~Sun, and
  N.~Thawdar, ``X60: A programmable testbed for wideband 60 ghz wlans with
  phased arrays,'' \emph{Computer Communications}, vol. 133, pp. 77--88, 2019.
  [Online]. Available:
  \url{https://www.sciencedirect.com/science/article/pii/S0140366417312896}
\BIBentrySTDinterwordspacing

\bibitem{Ish_two_beams}
\BIBentryALTinterwordspacing
I.~K. Jain, R.~Subbaraman, and D.~Bharadia, ``Two beams are better than one:
  Towards reliable and high throughput mmwave links,'' in \emph{Proceedings of
  the 2021 ACM SIGCOMM 2021 Conference}, ser. SIGCOMM '21.\hskip 1em plus 0.5em
  minus 0.4em\relax New York, NY, USA: Association for Computing Machinery,
  2021, p. 488–502. [Online]. Available:
  \url{https://doi.org/10.1145/3452296.3472924}
\BIBentrySTDinterwordspacing

\bibitem{Agilelink}
\BIBentryALTinterwordspacing
H.~Hassanieh, O.~Abari, M.~Rodriguez, M.~Abdelghany, D.~Katabi, and P.~Indyk,
  ``Fast millimeter wave beam alignment,'' in \emph{Proceedings of the 2018
  Conference of the ACM Special Interest Group on Data Communication}, ser.
  SIGCOMM ’18.\hskip 1em plus 0.5em minus 0.4em\relax New York, NY, USA:
  Association for Computing Machinery, 2018, p. 432–445. [Online]. Available:
  \url{https://doi.org/10.1145/3230543.3230581}
\BIBentrySTDinterwordspacing

\bibitem{FALP}
N.~J. Myers, A.~Mezghani, and R.~W. Heath, ``Falp: Fast beam alignment in
  mmwave systems with low-resolution phase shifters,'' \emph{IEEE Transactions
  on Communications}, vol.~67, no.~12, pp. 8739--8753, 2019.

\bibitem{beamhistory}
\BIBentryALTinterwordspacing
L.~Patra, Avishek and P.~M\"{a}h\"{o}nen, ``Smart mm-wave beam steering
  algorithm for fast link re-establishment under node mobility in 60 ghz indoor
  wlans,'' in \emph{Proceedings of the 13th ACM International Symposium on
  Mobility Management and Wireless Access}, ser. MobiWac ’15.\hskip 1em plus
  0.5em minus 0.4em\relax New York, NY, USA: Association for Computing
  Machinery, 2015, p. 53–62. [Online]. Available:
  \url{https://doi.org/10.1145/2810362.2810363}
\BIBentrySTDinterwordspacing

\bibitem{pedestrians}
G.~R. MacCartney, T.~S. Rappaport, and S.~Rangan, ``Rapid fading due to human
  blockage in pedestrian crowds at 5g millimeter-wave frequencies,'' in
  \emph{GLOBECOM 2017 - 2017 IEEE Global Communications Conference}, 2017, pp.
  1--7.

\end{thebibliography}
